\documentstyle[sprocl,epsfig]{article}
\def\Journal#1#2#3#4{{#1} {\bf #2}, #3 (#4)}
\def\Journall#1#2#3#4#5{{#1} #2 {\bf #3}, #4 (#5)}

\def\NPA{{\em Nucl. Phys.} A}

\def\PLB{{\em Phys. Lett.}  B}
\def\IJMP{{\em Int. J. Mod. Phys.} A}

\def\PRD{{\em Phys. Rev.} D}
\def\PRC{{\em Phys. Rev.} C}
\def\ZPC{{\em Z. Phys.} C}
\def\PTE{\em Prib. Tekhn. Eksp.}

\def\ib{\em ibid.}
\def\ea{\em et al.}

\def\ve{\varepsilon}
\def\vp{\varphi_q}

\def\la{\lambda_q}

\def\be{\begin{equation}}
\def\ee{\end{equation}}
\def\bea{\begin{eqnarray}}
\def\eea{\end{eqnarray}}
\def\ifmath#1{\relax\ifmmode #1\else $#1$\fi}%

\def\cpq{C_{p,q}}

\def\ifmath#1{\relax\ifmmode #1\else $#1$\fi}%
\def\rd{\ifmath{{\mathrm{d}}}}

\def\rT{\ifmath{{\mathrm{T}}}}

\def\vs{\vskip}

\begin{document}

\pagestyle{empty}

\begin{center}
{\large \bf STUDY OF FRACTALITY AND CHAOTICITY 
IN CENTRAL 4.5$A$ GeV/$c$ C-CU COLLISIONS}
\bigskip
\bigskip

L.K. GELOVANI,  G.L. GOGIBERIDZE \\
\smallskip

{\it Joint Institute for Nuclear Research, P.O.B. 79,\\
Moscow 101000, Russia}
\medskip
\medskip

E.K.G. SARKISYAN \\
\smallskip

{\it School of Physics \& Astronomy,
Tel Aviv University,\\ Tel Aviv 69978, Israel}
\vspace*{1.5cm}

ABSTRACT
\end{center}
\medskip

{\small 
Multifractality of charged particle pseudorapidity distributions is
analyzed in central collisions of carbon and copper nuclei at 4.5 GeV/c
per nucleon.  Within the method of normalized factorial moments, modified
to remove the bias of infinite statistics in the normalization,
intermittency-like behavior (scaling) is observed up to eight-particle
fluctuations. The study indicates a possible non-thermal phase transition
inside cascading and two different regimes presented in multiparticle
production.  Dynamics of fractal structure formation is further studied
using the chaoticity approach recently proposed. The distributions of the
horizontal factorial moments are considered and a scaling behavior,
referred to as erraticity, of the normalized moments of the distributions
is obtained. The corresponding entropy indices are calculated indicating
chaotic nature of multiparticle production with a specific self-similar
structure. 
}
\vspace*{1.4cm}

\begin{center}
Talk given at

the 8th Symposium of the Hellenic Nuclear Physics Society

``Advances in Nuclear Physics and Related Areas''

Thessaloniki, Greece, July 8 -- 12, 1997 
\end{center}


\newpage
\pagestyle{plain}
\setcounter{page}{1}


\title{STUDY OF FRACTALITY AND CHAOTICITY \\
IN CENTRAL 4.5$A$ GeV/$c$ C-CU COLLISIONS}

\author{ \underline{L.K. GELOVANI},$^{1,a}$  G.L. GOGIBERIDZE,$^{1,}$\footnote
{On leave from  Institute of Physics, Tbilisi 380077, Georgia.}
E.K.G. SARKISYAN$\,^2$}
\smallskip
\address
{$^1$Joint Institute for Nuclear Research, P.O.B. 79,
Moscow 101000, Russia\\
$^2$School of Physics \& Astronomy,
Tel Aviv University, Tel Aviv 69978, Israel}


\maketitle\abstracts{
Multifractality of charged particle pseudorapidity distributions is
analyzed in central collisions of carbon and copper nuclei at 4.5 GeV/c
per nucleon.  Within the method of normalized factorial moments, modified
to remove the bias of infinite statistics in the normalization,
intermittency-like behavior (scaling) is observed up to eight-particle
fluctuations. The study indicates a possible non-thermal phase transition
inside cascading and two different regimes presented in multiparticle
production.  Dynamics of fractal structure formation is further studied
using the chaoticity approach recently proposed. The distributions of the
horizontal factorial moments are considered and a scaling behavior,
referred to as erraticity, of the normalized moments of the distributions
is obtained. The corresponding entropy indices are calculated indicating
chaotic nature of multiparticle production with a specific self-similar
structure.  
}


Study of geometrical structure of dynamical fluctuations produced in
high-energy collisions gives an exceptional opportunity to investigate
hadroproduction process.\cite{rev1} The intermittency/fractal patterns of
such density fluctuations observed in all types of collisions, one
connects with two possible scenarios.  A geometrical monofractal structure
is associated~\cite{bnp} with a (most expected) thermal phase transition
e.g., from a quark-gluon plasma to nuclear matter, while multifractality
is assigned~\cite{pesch} to a self-similar cascading from partonic state
to final hadrons with a possible ``non-thermal'' (non-equilibrium) phase
transition during the cascade.  On the other hand, the quantities under
the study are usually calculated through averaged moments of the
distributions and changes of the density fluctuations from event to event
are not taken into account.  These lead to the loss of information about
more structure, namely, about chaoticity nature of multiparticle
production. Recently, erraticity approach has been proposed to study
chaotisity in high-energy physics,\cite{hwa:er} in particular, to
calculate entropy indexes,\cite{hwa:en} large value of which can be
considered as a signature of chaotic dynamics in self-similar
multiparticle production (e.g.  in branching processes in
QCD~\cite{hwa:enq}).

In this report we perform an investigation of dynamics of fractality and
chaoticity in multihadron production in central collisions of relativistic
nuclei. Such studies have a specific interest due to the above mentioned
expectation of quark-gluon plasma formation in high-energy nuclear
reactions and, on the other hand, in a sense of a possible hadronic nature
of intermittency, being also stronger with energy decrease.\cite{rev1}
Note, the study of fractality continues our previous
investigations,\cite{my3,my4} where different regimes in multiparticle
productions has been mentioned.

The contribution presented deals with the data came from interactions of
the JINR Synchrophasotron (Dubna)  4.5~$A$ GeV$/c$ $^{12}$C beam with a
copper target inside the 2m Streamer Chamber SKM-200.\cite{skm} A central
collision trigger was used: absence of charged particles with momenta
$p>3$ GeV$/c$ in a forward cone of 2.4$^{\circ}$ was required. 

The scanning and the handling of the film data were carried out on special
scanning tables of the Lebedev Physical Institute (Moscow).\cite{obr} The
average measurement error in the momentum $\langle\ve_p/p\rangle$ was
about 12$\%$, and that in the polar angle measurements was
$\langle\ve_{\vartheta}\rangle\simeq 2^{\circ}$. In total, 663 events with
charged particles in the pseudorapidity window $\Delta \eta=0.2-3.0$ are
considered ($\eta=-\ln\tan (\vartheta/2)$). The accuracy
$\langle\ve_{\eta}\rangle$ does not exceed 0.1. In addition, particles
with $p_\rT>1$ GeV/$c$ are excluded from the investigation as far as no
negative charged particles were observed with such a transverse momentum. 
Under the assumption of an equal number of positive and negative pions,
this cut was applied to eliminate the contribution of protons. The average
multiplicity is $23.8\pm 0.4$.

The density fluctuations are considered in the $\eta$-phase-space. To
avoid the problem connected with a non-flat shape of the distribution
$\rho (\eta)$, we use the ``cumulative'' variable,\cite{fl}

\begin{equation}
\stackrel{\sim }{\eta}\; =
\int_{\eta_{min}}^{\eta}
\rho(\eta')\rd \eta' \,
/
\int_{\eta_{min}}^{\eta_{max}} \rho(\eta')\rd \eta'\; ,
\label{nv}
\end{equation}
with  the
uniform spectrum $\rho(\stackrel{\sim }{\eta})$  within the interval [0,1].

Self-similar (fractal) dynamics is revealed by a scaling-law,

\be
 F_q\, \propto \, M^{\vp}\ ,
\qquad 0<\vp\leq q-1\  , \ 
\qquad (q\geq 2)\, ,
\label{fi}
\ee
of the $q$th order normalized (``vertical'') factorial  
moments,\cite{mm,blaz:mm}

\be
F_q =
\frac{{\cal N}^q}{M}
\sum_{m=1}^{M}
\frac{\langle n_m^{[q]}\rangle }
{N_m^{[q]}
}
\  ,
\label{fb}
\ee
over $M$ number of the bins into which the phase space of produced
particles is divided. Here, $n_m$ is the number of particles in the $m$-th
bin per event, $N_m$ is the similar number but calculated for all $\cal N$
events, and $\langle \cdots \rangle $ denotes averaging over the events. 
$n^{[q]}$ is the $q$th power factorial multinomial, $n(n-1)\cdots
(n-q+1)$. Note, we apply the modified moments proposed~\cite{mm} to remove
the biased estimator of the normalization (especially in small bins), that
along with the use of the ``transformed'' variable (\ref{nv}) allow to
study higher-order moments.

\begin{center}
\vs 2.mm
\begin{tabular}{lr}
\epsfig{file=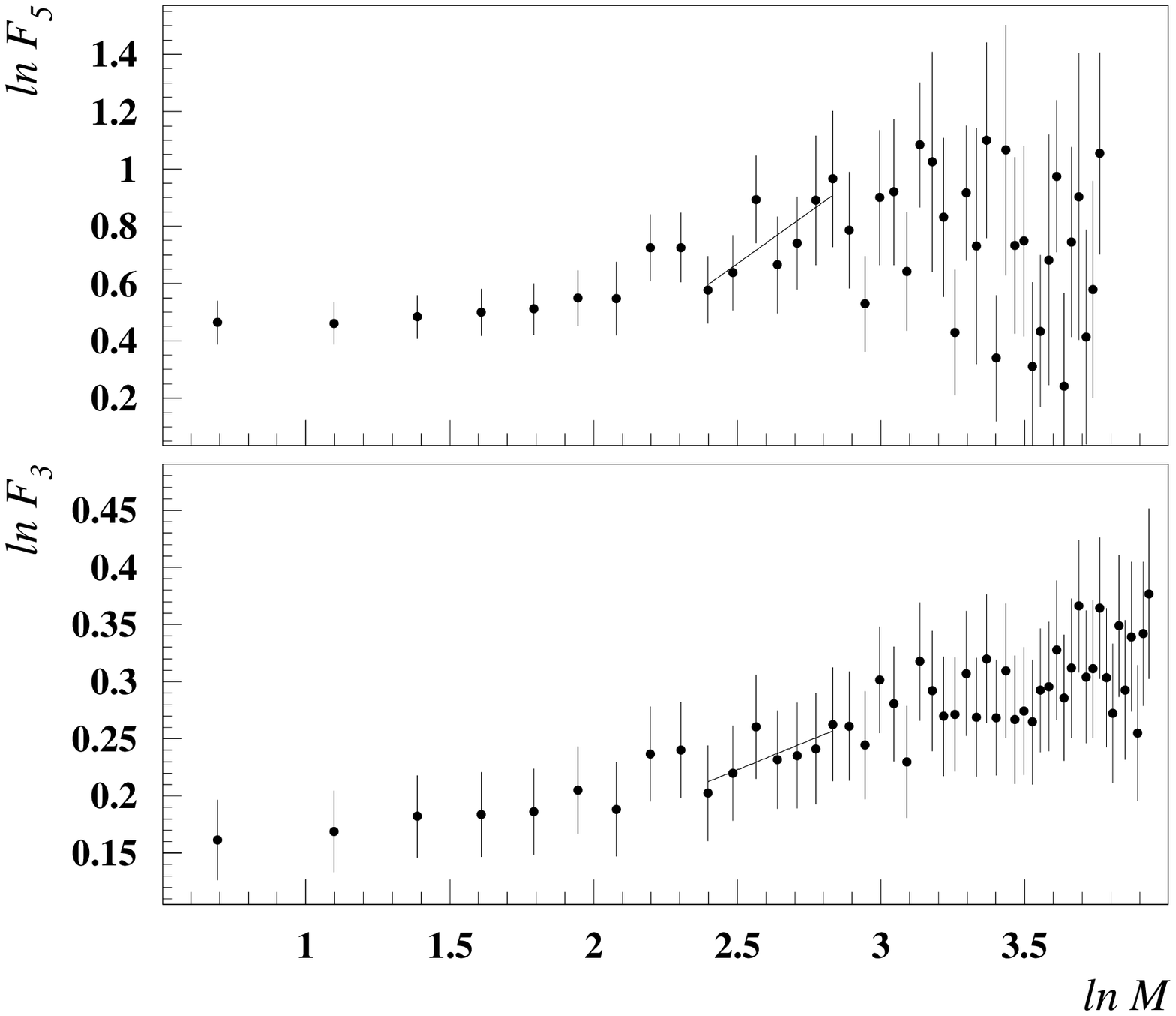,height=2.6in}
&
\epsfig{file=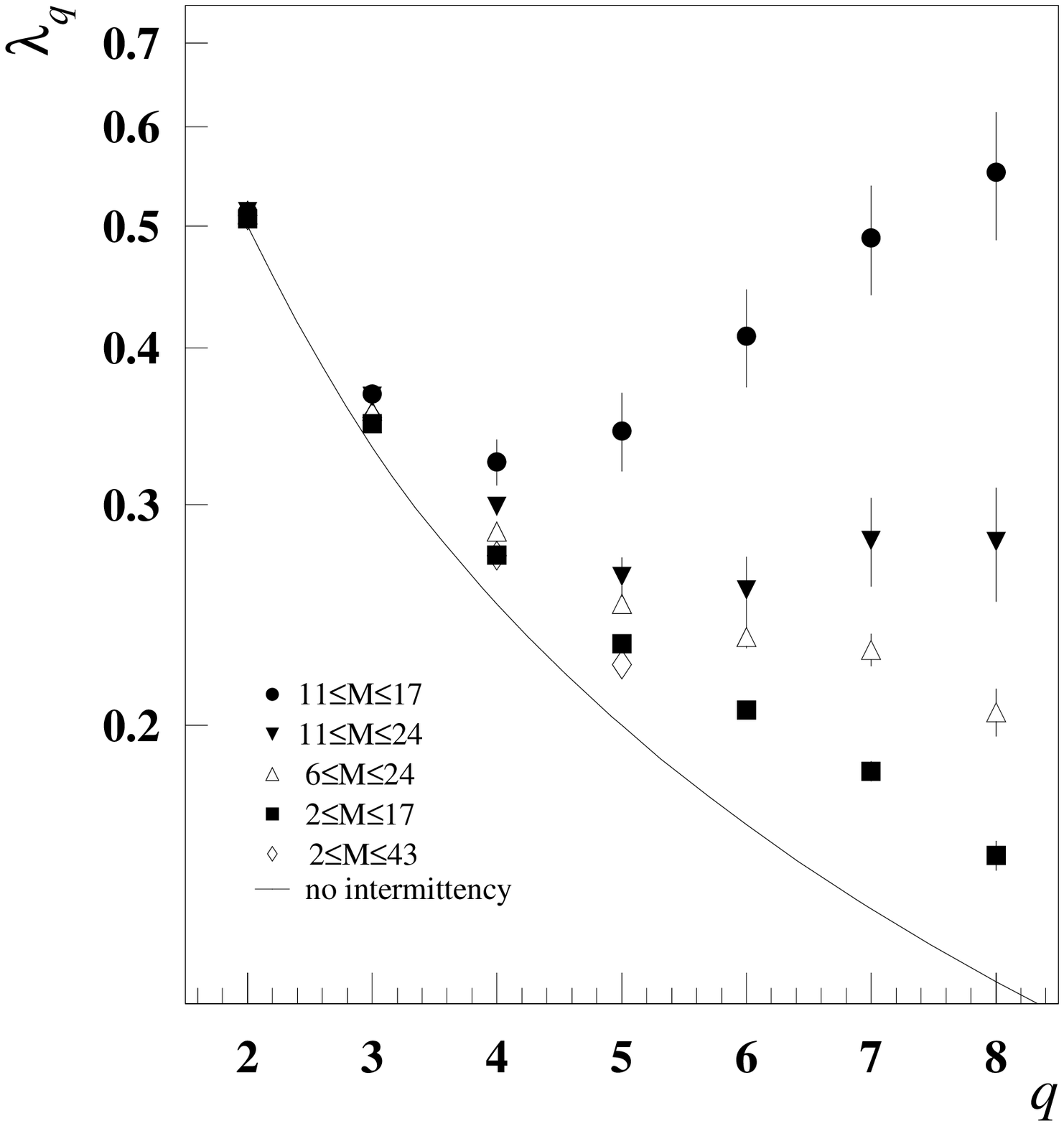,height=2.6in}
\end{tabular}
\end{center}
\vs 0.3cm
\centerline{\footnotesize{Fig. 1: $\ln F_q$ (\ref{fb})
vs. $\ln M$ for $q=3$ and 5.~~~~~~~~~~~~~~~~~~~~~~~~Fig. 2: $\la$
(\ref{lam}) vs. $q$.}}
\vs 0.4cm

The exponents $\vp$ show the strength of an intermittent structure of the
distribution. Fractality is characterized by the codimensions,
$d_q=\vp/(q-1)$, indicating monofractal objects if $d_q ={\rm const.}(q)$,
and multifractal ones if the chierarhy, $d_q>d_p$, $q>p$, is satisfied.
Then, a corresponding connection with a phase transition can be found as
discussed in the beginning.

As a signal of the transition, the existence of a minimum of the function
\be
\la=(\vp+1)/q
\label{lam}
\ee
at a certain ``critical'' value of $q=q_c$ is suggested.\cite{pesch} The
minimum of eq. (\ref{lam}) may also be a manifestation of coexistence of
many small (liquid-type) fluctuations and a few high-density
ones.\cite{bnp}

Fig.~1 presents dependence of $\ln F_q$ on $\ln M$, depicted for $q=3$ and
$5$. The different increase of the moments with increasing $M$ for the
different $M$-regions seeing on the plots continues up to $q=8$ (not
shown), confirming our earlier results~\cite{my3,my4} of the existence of
distinguished regimes of particle production at various bin-averaging
scales.

As in the preceding analysis, multifractality in hadroproduction has been
observed, i.e. $d_q>d_p$ at $q>p$ (not shown), supporting a cascading
scenario of particle production. Different increase of the $d_q$ with $q$
was found, indicating possible change of the regime of particle creation.

Fig.~2 shows the $\la$ function (\ref{lam}), confirming that at least two
regimes of particle production exist:  one with the phase transition at
$q_c$ between 4 and 5, and another one for which no critical behavior is
reached. The $q_c$-value and the $M$-intervals, which exhibit the minimum
of $\la$, $11 \leq M \leq 17$, $11\leq M\leq 24$, are found to be about
the same as in our previous studies~\cite{my3,my4} as well as in recent
similar investigations of heavy-ion collisions at ultra-high
energies.\cite{uhc} The most ``critical'' $M$-region, $11\leq M\leq 17$,
is shown by linear fit in Fig.~1.

Taking into account multifractality, the critical $q_c$
indicates~\cite{pesch} a ``non-thermal'' phase transition rather during
the cascade than within one phase.  Although the interpretation may be a
matter of debate, it must be noted that the minimum was found earlier also
in hadronic interactions~\cite{rev1} at small $p_\rT$ and has been
indicated in high-energy nuclear interactions.\cite{j}

\begin{center}
\vs 3.5mm
\begin{tabular}{lr}
\epsfig{file=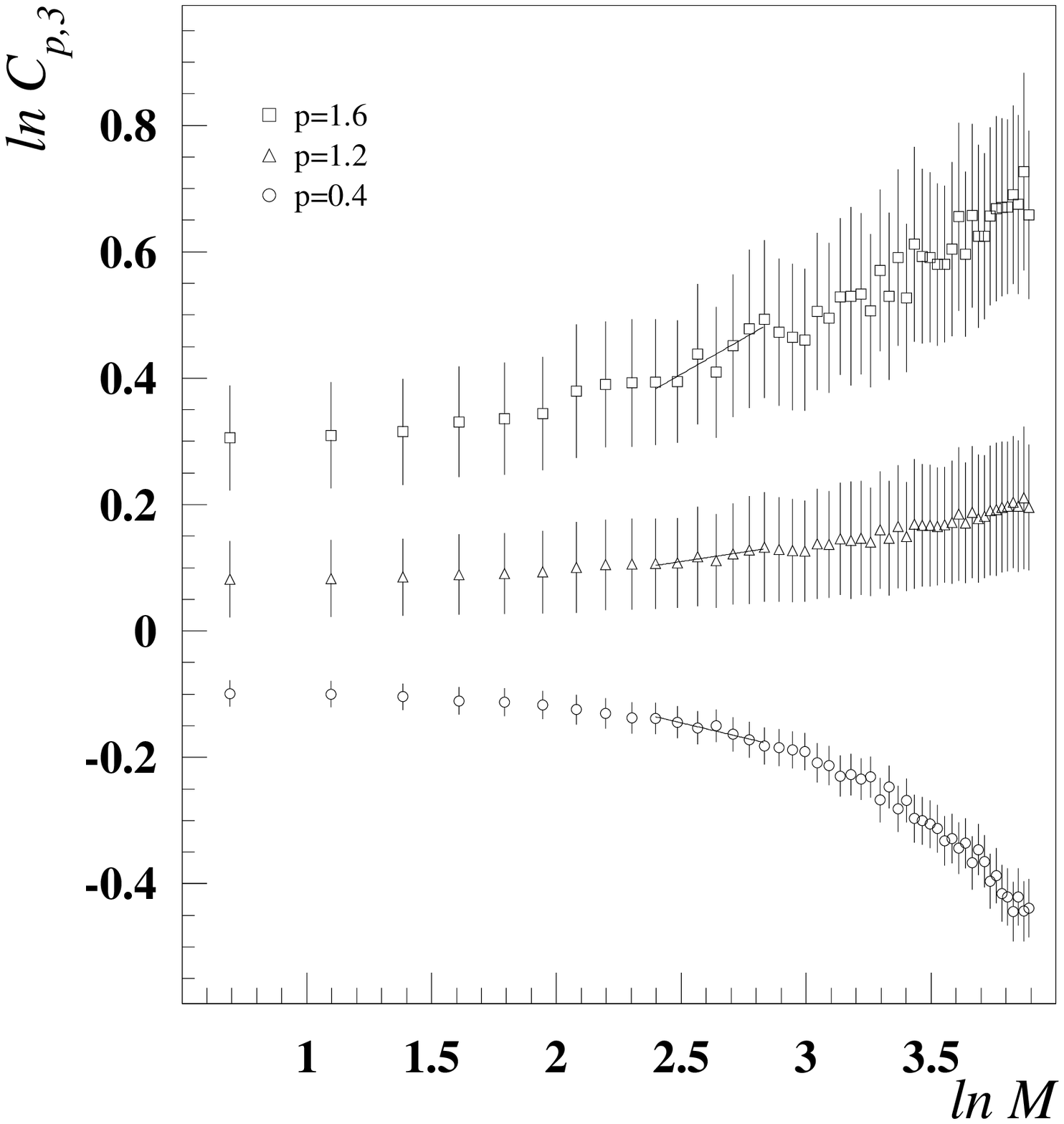,height=2.6in}
&
\epsfig{file=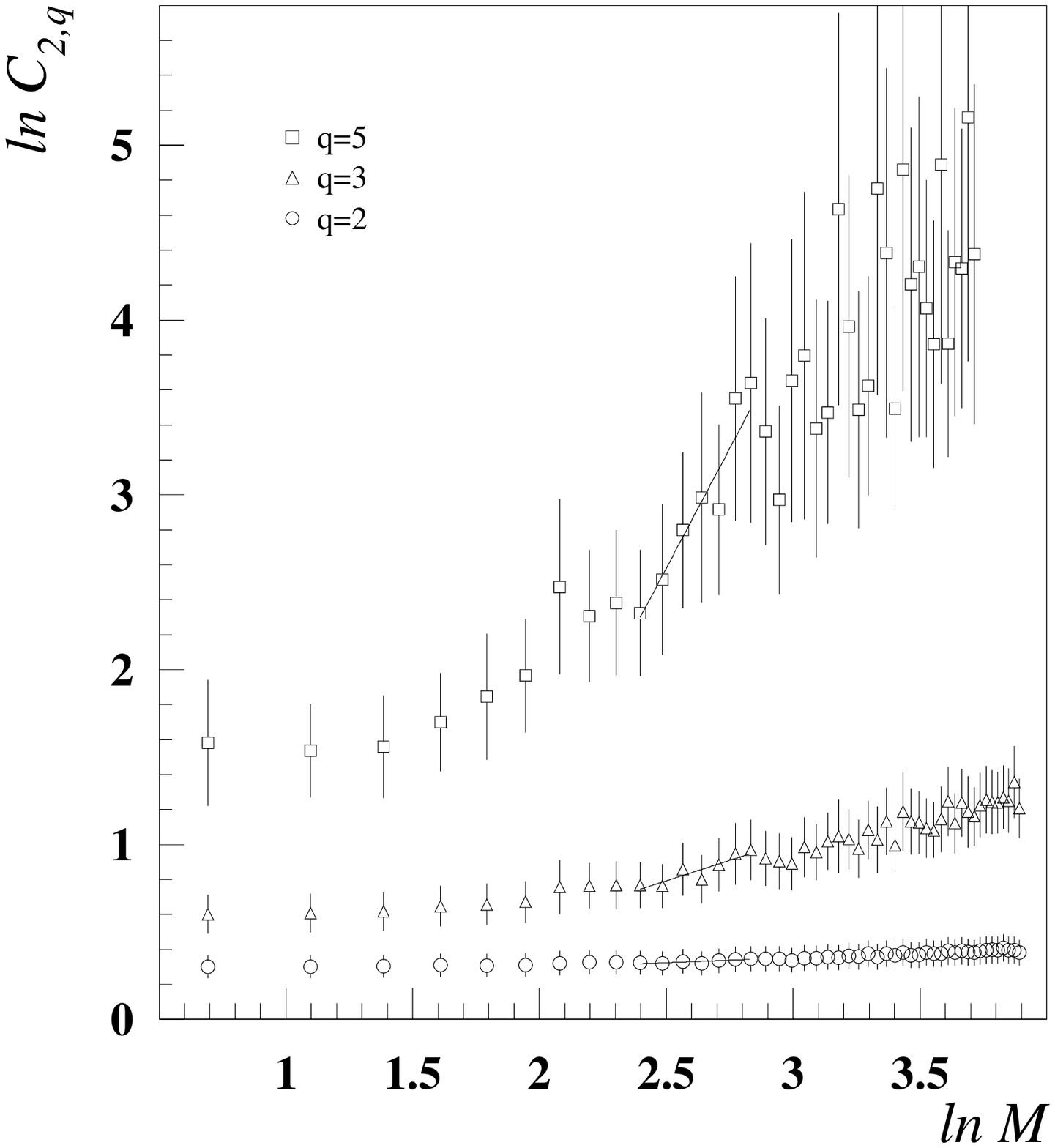,height=2.6in}
\end{tabular}
\end{center}
\vs 0.3cm
\centerline{\footnotesize{Fig. 3: $\ln C_{p,3}$ (\ref{cpq})
vs. $\ln M$.~~~~~~~~~~~~~~~~~~~~~~~~~~~Fig. 4:
$\ln C_{2,q}$ (\ref{cpq}) vs. $\ln M$.}}
\vs 0.4cm

Intermittency has not fully exhausted the charachteristics of dynamical
fluctuations. To study change of such fluctuations from  event to event
and then, to find out the chaoticity characteristics of the observed
self-similar process it was proposed~\cite{hwa:er} to consider the
normalized moments $\cpq$, defined by
\be
\cpq =\langle f_q^p \rangle /(F_q^H)^p\ ,
\label{cpq}
\ee
of the distributions of the normalized ``horizontal'' factorial 
moments~\cite{mm}, $F_q^H=\langle f_q \rangle$. Here, 
\be
f_q=
\frac{{\cal N}^q}{M}
\sum_{m=1}^{M}
\frac{n_m^{[q]}}
{(N/M)^{[q]}}
\label{fh}
\ee
and $N$ is a total number of particles. Note, the order $p$ of the
$\cpq$-moments can be non-integer, negative or zero, but are restricted
with zero $f_q$'s (``empty-bin effect'').

The degree of erratic fluctuations of the self-similar dynamics, indicated
by (\ref{fi}), is suggested to have a scaling behavior,
\be
\cpq \, \propto \, M^{\psi_{p,q}}\, \, .
\label{psi}
\ee
The exponents $\psi_{p,q}$ are referred to as erraticity indices and show a
strength of ``spatial'' fluctuations.

The power-law~(\ref{psi}) of the $\cpq$-moments vs. $M$ is shown in
Figs.~3 and 4. The first one illustrates the $M$-dependence of the moments
for different $p=0.4, 1.2,$ and $1.6$, and fixed $q=3$, while Fig.~4
presents the case, when $p$ is fixed, $p=2$, and $q$ changes, $q=2,3,5$.
As seen, the scaling behavior is satisfied up to small bins, however for
$q>2$ contributions of zero $f$'s (\ref{fh}) are sensitive making the
moments to saturate. This also is a reason why we do not consider higher
$q$-orders ($q\geq 6$) as well as $p<0$.

Note, shown by linear fit the most ``critical'' $M$-interval in Figs.~3
and 4 exhibits the same rise character as in Fig.~1. Corresponding
erraticity indexes are shown in Fig.~5a, compared to the ``non-critical''
fit for $2\leq M \leq 17$. The fourth order polynomials are used to
approximate the $\psi_{p,q}$-slopes as functions of $p$. The general
properties~\cite{hwa:er} of the $\psi$-functions are seen: $\psi_{p,q}<0$
for $0<p<1$, $\psi_{p,q}>0$ for $p>1$ (see also Fig.~3). Nevertheless, it
is clear, that the fits deviates from zero for $p=0$ that is connected
with the empty-bin contributions.

\begin{center}
\vs 6.mm
\begin{tabular}{lr}
\epsfig{file=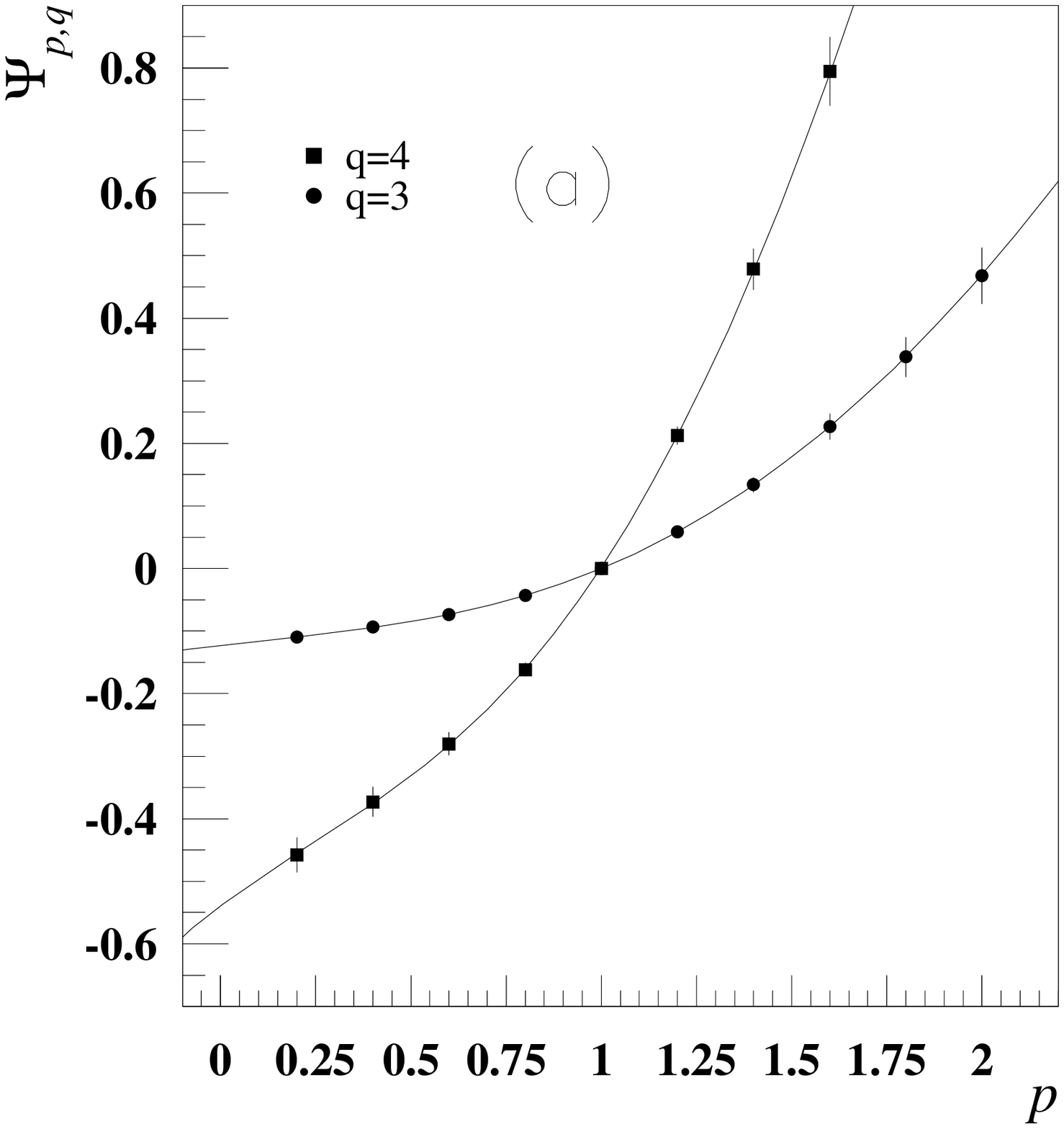,height=2.6in}
&
\epsfig{file=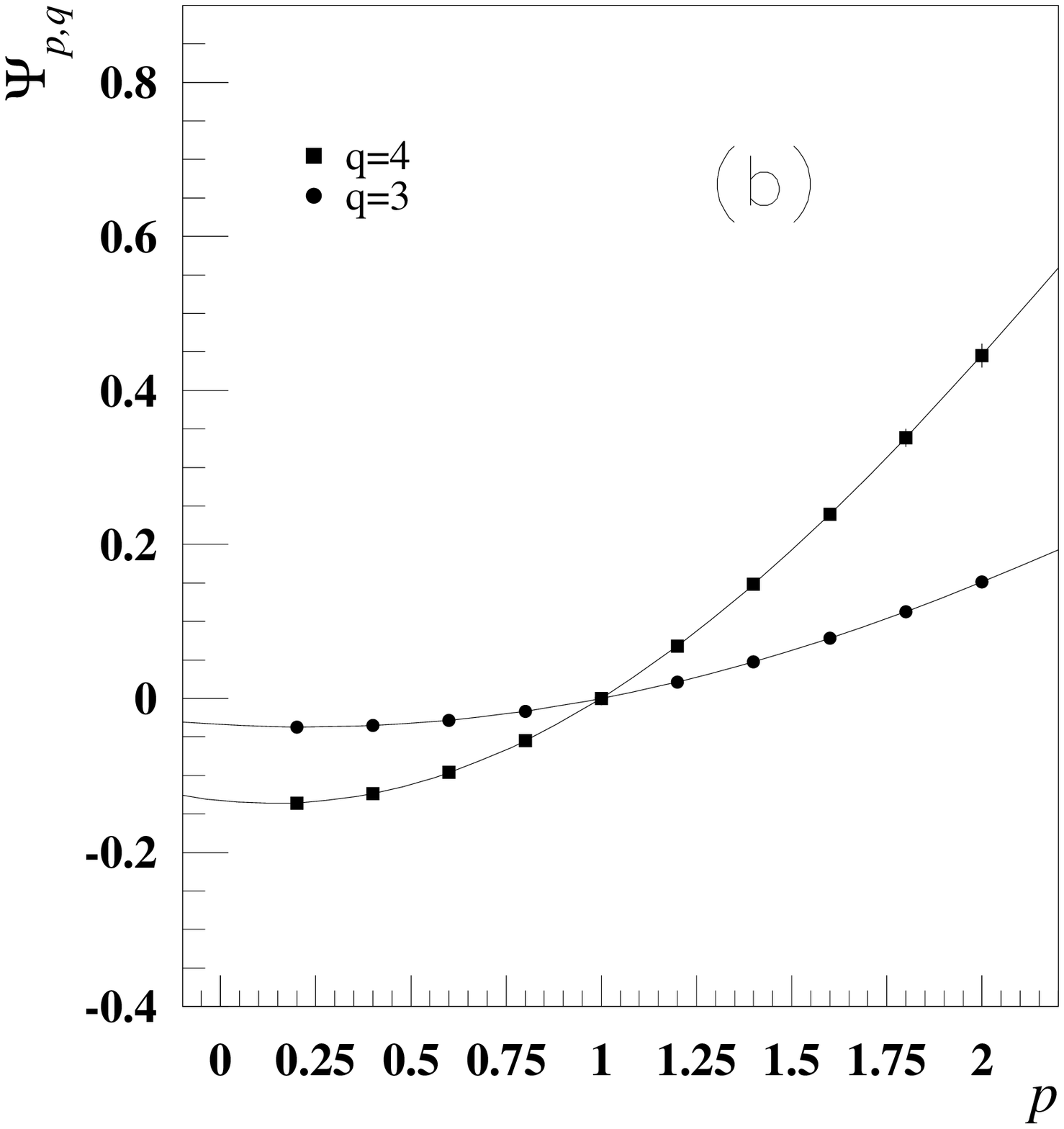,height=2.6in}
\end{tabular}
\end{center}
\vs 0.3cm
\begin{center}
{\footnotesize Fig. 5: Erraticity indexes $\psi_{p,q}$
(\ref{psi}), $q=3,4,$ as functions of $p$ for different $M$-intervals:\\
(a) $11\leq M\leq 17$;
(b) $2\leq M\leq 17$.}
\end{center}
\vs 0.4cm

Of particular interest is an index $\mu_q$ defined as

\be
\mu_q=\frac{\rd }{\rd p}\psi_{p,q}\big|_{p=1}\ ,
\label{mu}
\ee
shown~\cite{hwa:en,hwa:enq} to be related to the entropy in event space and be
a possible measure of chaoticity. The larger $\mu_q$ is, the more chaotic the
system is. In the case of a phase transition high values of $\mu_q$  are
expected.\cite{hwa:rep}

Table~1 presents the $\mu_q$ calculated for different $M$-intervals and
$q=2...5$. The large values found for each interval indicate very chaotic
dynamics of particle production, confirming its cascading nature. It is
worthwhile to note increase of the entropy index with approaching to the
``critical'' $M$-intervals that lends support for a phase transition. 
However, we must emphasize contribution of empty bins at high $q$'s.

\newpage
\begin{flushleft}
{\footnotesize Table 1. The entropy indexes $\mu_q$ (\ref{mu}).}
\end{flushleft}
\vs 0.1cm
\begin{center}
\begin{tabular}{c c c c c} \hline
\hline
$ M$-interval & $q=2$ & $q=3$ & $q=4$ & $q=5$ \\
\hline
11-17
&
$0.033\pm0.014$ & $0.251\pm0.081$ & $0.926\pm0.218$ & $2.070\pm0.355$
\\
11-24 &
$0.032\pm0.012$ & $0.222\pm0.023$ & $0.743\pm0.082$ &$1.581\pm0.170$
\\
2-17 &
$0.012\pm0.002$ & $0.095\pm0.010$ & $0.309\pm0.034$ & $0.631\pm0.065$
\\
2-43 &
$0.024\pm0.001$ & $0.175\pm0.005$ & $0.516\pm0.014$ & $0.967\pm0.025$
\\
\hline
\hline
\end{tabular}
\end{center}
\vs .2cm

In summary, the dynamics of fractality and chaoticity in multiparticle
production is studied with the pseudorapidity spectra of charged particles
produced in central C-Cu collisions at 4.5 GeV/$c$/nucleon. The scaled
factorial moments are calculated in the transformed variables and
corrected to take into account the bias of infinite statistics.
Multifractal structure of pseudorapidity density fluctuations is observed
to be of dynamical nature, indicating a possible non-thermal phase
transition. The existence of two different regimes of particle production
during the hadronization cascade is mentioned. These results based on the
increased statistics confirm conclusions of our previous
analysis.\cite{my3,my4} Changes of the dynamical fluctuations from event
to event are further investigated using the distributions of the
horizontal factorial moments. The normalized moments of these
distributions are studied and a scaling behavior referred to as erraticity
is obtained. The ``spatial'' entropy indexes are calculated shown strong
chaotic nature of the hadroproduction process, especially in the case of
possible non-thermal phase transition.

\section*{Acknowledgments}

L.G. is grateful to the Organizing Committee for inviting her to
participate in this Conference, supporting and kind hospitality. 
She is also thankful to the Russian Found for Fundamental Research for
supporting her work under grant 96-02-19359a. 
E.S. thanks Prof. R.C. Hwa for helpful discussions.

\end{document}
